\documentclass[12pt]{article}
\usepackage{amsmath}
\usepackage{amssymb}
\usepackage{amsthm}
\usepackage{graphicx}
\usepackage{psfrag}
\usepackage[hang, nooneline]{subfigure}
\usepackage{color}

\usepackage{soul} 
\usepackage{cite}

\usepackage{amssymb} 
\usepackage{amsfonts} 
\usepackage{amsmath} 
\usepackage{slashed}

\usepackage{egothic}%
\usepackage{pgothic}%
\usepackage{yfonts}%

 \usepackage{yhmath} 

\newcommand{\we}{\wedge}

\newcommand{\der}{\partial}

\newcommand{\inn}{\hspace*{2pt}\raisebox{-1pt}{\rule{6pt}{.3pt}\hspace*
{0pt}\rule{.3pt}{8pt}\hspace*{3pt}}}

\newcommand{\beq}{\begin{equation}}
\newcommand{\eeq}{\end{equation}}
\newcommand{\beqa}{\begin{eqnarray}}
\newcommand{\eeqa}{\end{eqnarray}}
\newcommand{\nn}{\nonumber}

\newcommand{\pbr}[2]{ \{ \hspace*{-2.6pt} [ #1 , #2\hspace*{1.4 pt} ] 
\hspace*{-2.6pt} \} }

\author{ 
\vspace*{1ex} \\
Monika E. Pietrzyk$^{1}$,  C\'ecile Barbachoux$^{2}$,  
   Igor V. Kanatchikov$^{3,4}$   \\ 
and Joseph Kouneiher$^{2}$ 
\\ \small $^{1}\!$ Mathematics and Physical Sciences, University of Exeter, 
 EX4 4QL Exeter, UK
\\ 
\small {\!\!\!}${\!\!\!}^{2}\!$  
Sciences and Technologies Department/INSPE, Universit\'e C\^ote d'Azur,
06000 Nice, France 
\\  
\small ${}^{3}$ National Quantum Information Centre KCIK, 81-931 Sopot, Poland 
\\
\small ${}^{4}$ IAS-Archimedes Project,  Côte d'Azur, France
}

\date{} 

\title{
\Large \bf 
On the covariant Hamilton-Jacobi formulation of Maxwell's equations 
 via the polysymplectic reduction 
}

\begin{document}

\maketitle

\begin{abstract}
The covariant Hamilton-Jacobi formulation of Maxwell's equations 
is derived from the first-order (Palatini-like) Lagrangian  using the analysis of constraints within the De~Donder-Weyl covariant Hamiltonian formalism and the corresponding polysymplectic reduction. 
\end{abstract}

\section{Introduction} 

Recently we have constructed the covariant Hamilton-Jacobi equation of the teleparallel equivalent of General Relativity in Palatini formulation \cite{mcec}.  The construction applies the generalized Dirac--Bergmann algorithm developed 
in \cite{ik-dirac} to the analysis of second-class constraints within the covariant Hamiltonian theory of De Donder and Weyl \cite{dedonder,weyl,kastrup,rund,vonrieth} (also known as polysymplectic, finite-dimensional, multisymplectic, and other names). The analysis relies on 
 the generalization of Poisson brackets to the De Donder--Weyl (DDW) extension of the Hamiltonian 
 formulation to field theories \cite{ikbr1,ikbr2,ikbr3,ik5,khbr1,khbr2,khbr3}.

It is well known that the Hamilton-Jacobi formulation of mechanics is the classical or geometric optics approximation of the quantum Schr\"odinger wave equation, and that the arguments based on the Hamilton-Jacobi theory have led to the discovery of Schr\"odinger's famous equation \cite{butterfield} and to the interpretation of quantum mechanics known as ``Bohmian mechanics" 
or ``causal interpretation" \cite{hiley,holland,durr,bhk87}. The canonical Hamilton-Jacobi formulation of General Relativity \cite{peres2} played a heuristic role in the formulation of the famous Wheeler-DeWitt equation in canonical quantum gravity \cite{kiefer}. 

In the context of field theory, we usually think that the Hamilton-Jacobi formulation is 
an infinite-dimensional generalization of the Hamilton-Jacobi formulation in mechanics 
 in which the configuration space is an infinite-dimensional space of field configuration and the variational derivatives replace partial derivatives \cite{peres2,frechet,volterra-book}. 
However, two different approaches to the generalization of the Hamilton-Jacobi theory to field theories, which are obtained from multidimentional variational problems, existed from the very beginning in the pioneering paper by Vito Volterra \cite{volterra}. The infinite-dimensional variational derivatives path has been taken by the canonical formalism in field theory \cite{weiss},  and the finite dimensional partial derivatives path   
is better known in the geometric calculus of variations 
\cite{dedonder,weyl,vonrieth,rund,kastrup}. 

The covariant Hamilton-Jacobi theory of De Donder and Weyl  \cite{dedonder,weyl,vonrieth,kastrup}, which treats the space and time variables equally,  has inspired the development of precanonical quantization of fields in \cite{ik1,ik2,ik3,ik4,ik5,ik5e}, its application in quantum gauge theory 
\cite{iky1,iky2,iky3} 
and quantum general relativity both in metric variables \cite{ikm1,ikm2,ikm3,ikm4} and vielbein variables \cite{ikv1,ikv2,ikv3,ikv4,ikv5}. Different attempts to use a manifestly covariant Hamilton-Jacobi formulation as a 
basis for the quantization of fields  can be found in \cite{schmidt,cremas}. 
Our earlier work \cite{mcec} was inspired by the recent application of precanonical quantization to quantization of the  teleparallel equivalent of general relativity \cite{ik-mg21,ik-tpq}.

Both the covariant Hamilton-Jacobi theories and precanonical quantization treat spacetime variables  equally and, consequently,  describe classical and quantum fields by means of functions on a finite-dimensional configuration space of spacetime variables and field variables. The former enter as a multidimensional analog of one-dimensional time in mechanics, and the latter is similar to generalized coordinates of particles in mechanics. The naturally arising questions of the relationship between this finite-dimensional description and the canonical description referring to the infinite-dimensional 
spaces of configurations of fields have been addressed in \cite{ik-pla,riahi} for the 
respective Hamilton-Jacobi formulations and in \cite{ik-pla,iks1,iks2,iksc1,iksc2,iksc3,iky1,iky3} for the precanonical quantization vs. the Schr\"odinger functional picture in QFT. 

In this paper we would like to derive the covariant De Donder--Weyl Hamilton-Jacobi equation for Maxwell's field using the approach of our previous paper \cite{mcec}. One motivation is to test the method of analysis of constraints within the De Donder--Weyl Hamiltonian formalism. The second motivation is to obtain the covariant Hamilton-Jacobi equation for the electromagnetic field which 
can serve as a proper equation for numerical integration near caustics 
and as a verification tool of the consistency  of precanonical quantization of gauge fields with the correct classical limit. 

We will proceed as follows. In section 2 we present the Palatini-like Lagrangean formulation 
of Maxwell's field. In section 3 we analyze the De Donder--Weyl (DDW) Hamiltonian formulation and 
conclude that the Palatini-like Lagrangian leads to a singular DDW Hamiltonian system with second class constraints. In section 4 we evaluate generalized Dirac brackets  and show how the constrained DDW Hamiltonian formulation of the previous section reduces to the unconstrained DDW Hamiltonian formulation on the reduced polymomentum phase space equipped with the corresponding reduced polysymplectic structure. The knowledge of the correct polysymplectic reduction allows us to write down the covariant DDW Hamilton-Jacobi equation in section 5 based on the previous geometric formulations of Hamilton-Jacobi theories in the multi-, poly-, and k-symplectic contexts. 
In section 6 we present conclusions and perspectives of further research.

\section{The first order formulation}

Instead of the usual Lagrangian of Maxwell's theory 
\beq\label{e1}
L = -\frac14 F_{\mu\nu}F^{\mu\nu},
\eeq
where 
\beq\label{e2}
F_{\mu\nu} = \der_\mu A_\nu - \der_\nu A_\mu,
\eeq 
let us start with a Palatini-like Lagrangian 
\beq \label{paction}
L = \frac14 P^{\mu\nu} ( P_{\mu\nu} - 2 F_{\mu\nu} ),
\eeq 
where the electromagnetic potential $A_\mu$ and the antisymmetric quantities 
 $P_{\mu\nu} = - P_{\nu\mu}$  are independent field variables. 

The variation of $P_{\mu\nu}$ leads to 
\beq  \label{e3}
P_{\mu\nu} =  F_{\mu\nu} 
\eeq
and the variation of $A_\mu$ leads to 
\beq \label{e4}
\der_\mu P^{\mu\nu} = 0.
\eeq
 The substitution of (\ref{e3}) in (\ref{e4}) reproduces the Maxwell equations in the standard form 
$\der_\mu F^{\mu\nu}\!=\!0$ which is usually derived from the Lagrangian (\ref{e1}). 

In this paper we chose the Palatini-like formulation because it is the closest to the formalism used 
in our previous paper \cite{mcec} and the main motivation of the present consideration is a demonstration that the first-order formulation and the analysis of the appearing constraints using the procedure based on the brackets of forms and generalized Dirac brackets leads to the correct (known) result for DDW Hamilton-Jacobi equation for the vacuum electromagnetic field \cite{kastrup,vonrieth,horava}. 

\section{The De~Donder--Weyl Hamiltonian formulation}

Using the procedure of the De Donder--Weyl (DDW) Hamiltonian formulation we 
first obtain the expressions of the polymomenta associated with the independent field variables 
$A_\mu, P^{\mu\nu}$ 
\begin{align}
\label{e6}
p^\mu_{A_\nu} &= \frac{\der L}{\der \der_\mu A_\nu} =  - P^{\mu\nu}, \\
\label{e7}
p^\alpha_{P^{\mu\nu} } &= \frac{\der L}{\der \der_\alpha  P^{\mu\nu} } = 0.
\end{align} 

This results in a constrained DDW Hamiltonian system with the constraints 
\begin{align}
\label{e9}
C^\mu_{A_\nu}  &= p^\mu_{A_\nu} + P^{\mu\nu} \approx 0, \\
\label{e10}
C^\alpha_{P^{\mu\nu} } &= p^\alpha_{P^{\mu\nu} }   \approx 0.
\end{align}
Note that a consequence of (\ref{e9}) 
\beq
\label{e6s}
p^\mu_{A_\nu} + p^\nu_{A_\mu} \approx 0 . 
\eeq
We use the standard Dirac's notation $\approx$ for weak equalities on the surface of constraints 
defined by (\ref{e9}) and (\ref{e10}).

The De~Donder--Weyl Hamiltonian function  has the form
\begin{align}
\label{e11}
H &= p^\mu_{A_\nu} \der_\mu {A_\nu} + p^\alpha_{P^{\mu\nu} } \der_\alpha P^{\mu\nu} - L 
\approx  - \frac12 P^{\mu\nu} F_{\mu\nu}
- \frac14 P^{\mu\nu} ( P_{\mu\nu} - 2 F_{\mu\nu} )
\nn \\ 
&= - \frac14 P^{\mu\nu} P_{\mu\nu} \approx - \frac14 p^\mu_{A_\nu}  p_\mu^{A_\nu}.
\end{align}

Following the generalization of the Dirac--Bergmann analysis of constraints to the 
DDW Hamiltonian systems \cite{ik-dirac} 
and using the notation 
\beq 
\upsilon_\alpha = \frac{\der}{\der x^\alpha} \inn\! (dx^0\we dx^1\we...\we dx^{n-1}) 
= (-1)^\alpha dx^0\we...\widehat{dx^\alpha} ... \we dx^{n-1} 
\eeq 
for the basis of $(n-1)-$forms in $n$-dimensions,
we introduce the $(n-1)-$forms of constraints  
\begin{align} \label{cforms}
\begin{split}
C_{A_\nu} &= C^\alpha_{A_\nu} \upsilon_\alpha,  \; \\
C_{P^{\mu\nu} } &= C^\alpha_{P^{\mu\nu}}\upsilon_\alpha
\end{split}
\end{align}
and calculate their brackets. 

The brackets of forms are defined by the polysymplectic 
 structure (first introduced in \cite{ik5,ikbr1,ikbr2,ikbr3}) 
 on the 
unconstrained polymomentum analog of phase space, the space of 
 field variables $(A,P)$  and their respective polymomenta ${p^\mu_A, p^\mu_P},$  
\beq
\Omega =  d A\we dp^\alpha_A \we\upsilon_\alpha + d P \we d p^\alpha_P \we \upsilon_\alpha.
\eeq
$\Omega$ maps a form $C$ of degree $(n-1)$ to a vector field $\chi_C$
\beq
\chi_C \inn \Omega = d C
\eeq 
and the bracket of two $(n-1)$-forms are defined as follows 
\beq 
\pbr{C_1}{C_2} := \chi_{C1}\inn d C_2. 
\eeq 
It is easy to see that the bracket of two $(n-1)$-forms is a $(n-1)$-form. 
Note that this definition produces brackets for a limited class of forms called Hamiltonian forms.  
 The Lie algebra structure defined by this bracket is embedded into a larger bi-graded 
 (Gerstenhaber algebra) structure defined on differential forms of all degrees from $0$ to $(n-1)$  
\cite{ik5,ikbr1,ikbr2,ikbr3,khbr1,khbr2,khbr3}. 

By a direct calculation, we obtain the following brackets of the forms of constraints 
\begin{align} 
\label{e14}
C_{{P^{\mu\nu}} {A_\sigma}} = \pbr{C_{P^{\mu\nu}}}{C_{A_\sigma}} &= 
\upsilon_{[\mu} \delta^{\sigma}_{\nu]},
\\
C_{AA'}=\pbr{C_A}{C_{A'}} &= 0, \\
C_{PP'}=\pbr{C_P}{C_{P'}} &= 0.
\end{align}
The right-hand side of (\ref{e14}) does not vanish on the surface of constraints. 
Hence we have here the DDW analog of the second-class constraints according to the Dirac classification \cite{dirac,henneaux,regge}. 

\section{The Dirac brackets}

A generalization of the 
Dirac bracket in the context of constrained DDW Hamiltonian systems with second-class constraints have been proposed in \cite{ik-dirac}. It has been used in precanonical quantization of Einstein gravity  \cite{ikv1,ikv2,ikv3,ikv4,ikv5}, in different models of gravity \cite{mex1,mex2,mex3} and gauge theories \cite{mex4}, 
our previous  work on the teleparallel equivalent of general relativity \cite{mcec} and its precanonical quantization in \cite{ik-tpq,ik-mg21}. 

For  $(n-1)$-- 
forms $A$ and $B$ 
\beq \label{pdir}
\pbr{A}{B}{}^* := \pbr{A}{B}{} - \sum\limits_{U,V}\pbr{A}{C_U}{}\bullet  C_{UV}^{\sim\! 1} \we \pbr{C_V}{B}{},
\eeq
where 
\beq
A\bullet B := *^{-1}(*A\we*B) , 
\eeq 
$*$ is the Hodge star,  
the indices $U,V$ enumerate the primary constraints, i.e., in the present case, they run over all the indices of 
$A_\mu$ and $P^{\mu\nu}$.  $C^{\sim 1} =C^{\sim\!1}_\alpha dx^\alpha $ denotes 
the pseudoinverse matrix defined by the relation 
\beq
C^{}\bullet C^{\sim 1} \we C^{} = C^{},
\eeq
which generalizes the definition of the Moore-Penrose pseudoinverse \cite{penrose} to the case of matrices whose components are exterior forms. Note that the distributive law for $\we$ and $\bullet$ products is that the wedge product $\we$ acts first. 

For the matrix of constraints 
\beq\label{matrc}
C_{UV}=
\begin{bmatrix}
   {C}_{AA'}=0    & \quad {C}_{AP} & \\
 - ({C}_{AP})^T     & \quad {C}_{PP'} =0 
\end{bmatrix}, 
\eeq
we obtain the pseudoinverse matrix 
\beq\label{pseinv}
C^{\sim\!1}_{UV}=
\begin{bmatrix}
 {C}^{\sim 1}_{AA'}=0    & \quad {C}^{\sim 1}_{AP} & \\
 - ({C}^{\sim 1}_{AP})^T     & \quad {C}^{\sim 1}_{PP'} =0    
\end{bmatrix}, 
\eeq
whose components satisfy the relations 
\begin{align}
\begin{split}
\sum_P C^\mu{}_{AP}C_{\mu PA'}^{\sim 1}&=\delta_{AA'}, \\
\sum_A C^\mu{}_{PA}C_{\mu AP'}^{\sim 1}&=\delta_{PP'}.
\end{split}
\end{align}

The calculation of generalized Dirac brackets leads to the result for brackets 
between $(n-1)-$forms of polymomenta $p_A = p_A^\alpha \upsilon_\alpha$, 
$p_P = p_P^\alpha \upsilon_\alpha$ and $(n-1)-$forms of field variables 
$A\upsilon_\alpha, P\upsilon_\alpha$ 
\begin{align}
\label{aa}
\pbr{p_A}{A'\upsilon_\alpha}^* &=  \pbr{p_A}{A'\upsilon_\alpha} = \delta_{AA'}\upsilon_\alpha, \\
\label{pa}
\pbr{p_P}{P'\upsilon_\alpha}^* &= \delta_{PP'}\upsilon_\alpha - 
\sum_{A,P''}\pbr{p_P}{C_A}\bullet C^{\sim\!1}_{AP''}\we\pbr{C_{P''}}{P'\upsilon_\alpha} = 0, \\ 
\label{ppa}
\pbr{P^{\mu\nu}\upsilon_\sigma}{A_\alpha\upsilon_\tau}{}^* &= 
- \sum_{P',A'}\pbr{P^{\mu\nu}\upsilon_\sigma}{C_{P'}}\bullet C^{\sim\!1}_{P'A'}\we\pbr{C_{A'}}{A_\alpha\upsilon_\tau } = C^{\sim\!1}_{\tau \, P^{\mu\nu} A_\alpha}  \upsilon_\sigma, \\
\pbr{p_P}{A\upsilon_\alpha}^* &= 0.
\end{align}

Hence the original polymomentum phase space of variables $(A,P, p_A, p_P)$ 
reduces to the space of $A$ and ${p_A}$ is equipped 
with the reduced polysymplectic structure given by 
\beq \label{omr}
\Omega_R = d A_{\mu} \we dp^\alpha_{A_{[\mu}} \we\upsilon_{\alpha ]},
\eeq
where the antisymmetrization appears as a consequence of the projection to the subspace 
of polymomenta which satisfies the constraint (\ref{e6s}). 

The De Donder--Weyl Hamiltonian on this reduced polysymplectic space assumes the form 
\beq
H^* = - \frac14  p^\alpha_{A_\mu} p_\alpha^{A_{\mu}}. 
\eeq

The DDW Hamiltonian field equations in terms of the Poisson bracket defined by the 
reduced polysymplectic structure takes the form 
\begin{align}
\label{e27}
d\bullet p^{\alpha}_{A_{[\mu }} \upsilon_{\alpha ]} &= 
\pbr{H^*}{p^{\alpha}_{A_{[\mu}} \upsilon_{\alpha ]} } = 
- \frac{\der H^*}{\der {A_\mu }} = 0, \\ 
\label{e28}
d\bullet A_{[\mu}  \upsilon_{\alpha ]} &= \pbr{H^*}{{A_{[\mu}}\upsilon_{\alpha ]}} \ = 
\ \frac{\der H^*}{\der{}{ p{}^{\, \alpha}_{ A_{\mu}}}}  
\, = -  \frac12 p_{ \alpha}^{A_{\mu}},
\end{align}
where the operation $d\bullet$ acts  on $(n-1)-$forms $N=N^\mu\upsilon_\mu$ as follows 
\beq
d\bullet N = \der_\alpha N^\mu dx^\alpha\bullet\upsilon_\mu = \der_\alpha N^\alpha.
\eeq

Hence, equation (\ref{e27}) reproduces the Maxwell equation in the form 
\beq
\der_\alpha p^\alpha_{A_\mu} = 0 
\eeq
and equation (\ref{e28}) reproduces the definition of the field strength represented by polymomenta 
$p_\alpha^{A_{\mu}} $ 
in terms of the vector potential (cf. (\ref{e6})) 
\beq 
\label{e34}
 \der_\alpha A_\mu -  \der_\mu A_\alpha  = - p_\alpha^{A_{\mu}}.
\eeq 
Note that the symmetric part of $p_\alpha^{A_{\mu}} $ is vanishing according to the constraint 
(\ref{e9}).

\section{The Covariant Hamilton-Jacobi equation} 

The geometrical formulation of the De Donder--Weyl Hamilton-Jacobi theory \cite{dlhj1,dlhj2,dlhj3,rr21} 
allows us to write the corresponding equation based on the reduced polysymplectic structure 
(\ref{omr}) derived in the previous section  
\begin{align}
\label{dwhj0}
\der_\mu S^\mu + H^* \left(p^\alpha_{A_{\mu}}=\frac{\der S^\alpha}{\der A_{\mu}}  , A_{\mu}\right) = 0.  
\end{align}

Therefore, the covariant De Donder--Weyl Hamilton-Jacobi equation for Maxwell's field takes the form 
\beq \label{e35}
\der_\mu S^\mu - \frac{1}{4} \frac{\der S^\mu}{\der A_\nu }\frac{\der S_\mu}{\der A^\nu }
=0.
\eeq
It is supplemented by the equation 
\beq \label{e36}
p^{(\mu }_{A_{\nu)}} = \frac{\der S^{(\mu}}{\der A_{\nu)} } 
=0,
\eeq 
which follows from the constraint (\ref{e6s})
and the ``embedding condition'' (cf. (\ref{e34})) 
\beq\label{e37}
p^{[\mu}_{A_{\nu ]}} = 
\frac{\der S^{[\mu}}{\der A_{\nu]} } 
= - F^{\mu\nu}.
\eeq
The result has been derived earlier in \cite{vonrieth,horava}  using other methods based on the Lagrangian (\ref{e1}). In this case, the only constraint is (\ref{e6s}) and the DDW Hamiltonian function is given by the last expression in (\ref{e11}). The covariant Hamilton-Jacobi equation can be written then in the form (\ref{dwhj0}) which reproduces solutions of the Maxwell equation when 
the functions $S^\mu$ satisfy the constraint (\ref{e36}) and the solutions are calculated from the solution of the Hamilton-Jacobi equation (\ref{e35}) using the embedding condition (\ref{e37}). 

Thus we have shown that the formalism of the analysis of constraints in the DDW theory based on a generalization of Poisson and Dirac brackets and a generalization of Dirac--Bergmann's analysis of constraints leads to the correct results in the well-understood and verifiable case of 
Maxwell's field. 
 
\section{Conclusions} 

The covariant DDW HJ equation is derived starting from the first-order Lagrangian formulation of Maxwell's equations. We intentionally started from a more singular Palatini-like formulation in order to test the method of dealing with singular DDW theories using the generalized Dirac bracket on differential forms introduced in \cite{ik-dirac}. Our procedure gives rise to a specific ``reduced polysymplectic structure" that allows us to use the existing geometrical formulations 
of the Hamilton-Jacobi theories in the context of closely related multisymplectic and $k$-symplectic formulations, both regular and singular in the sense of the DDW  Hamiltonian formalism. 
Note that the notion  of the ``polysymplectic reduction" in this paper is closer to the context of Dirac's theory of constraints generalized by Kanatchikov in \cite{ik-dirac},  and its relationship with other notions of the poly-/multisymplectic reduction and poly-/multisymplectic structures appearing in the mathematical literature (see e.g. \cite{rr-reduction,blacker,lopez} and the references therein) deserves a separate study. 


The DDW HJ equation (\ref{e35}) could be an interesting reformulation of Maxwell's equation from the point of view of numerical integration near singular configurations of electromagnetic fields.  Moreover, any quantization of the  electromagnetic field which is based on the DDW Hamiltonian formalism is required to reproduce this equation in the classical limit. In \cite{ik3,rutgers} the DDW HJ equation for the scalar field theory is obtained in the classical limit of the analogue of the Schr\"odinger equation put forward as the foundation of precanonical quantization \cite{ik3,ik4}.  It is interesting to understand how our DDW HJ equation (\ref{e35}) 
emerges in the classical limit of the precanonically quantized Maxwell field. The latter is easy to obtain from the more general case of Yang-Mills theories considered in \cite{ik5e,iky1,iky2,iky3}. 
Another question to ask is: how the partial derivative DDW HJ equation (\ref{e35})
is related to the canonical HJ equation for the electromagnetic field, which uses variational derivatives \cite{kaloy}?  
In \cite{ik-pla} and \cite{riahi} a similar question was discussed for the scalar field theory and gravity, respectively. The case of Maxwell's field will be considered in a follow-up  paper. 

Also, it is noteworthy that the previous work on different constrained DDW (polysymplectic) Hamiltonian systems \cite{mex1,mex2,mex3,mex4}  combined with the methods of this paper may allow us to obtain 
the covariant DDW HJ equations for these systems 
in a quite direct fashion.

\end{document}